# Accumulation of spin-polarized states of charge carriers and a spintronic battery

L.A. Pastur, V.V. Slavin, and A.V. Yanovsky

*B. Verkin Institute for Low Temperature Physics and Engineering of the National Academy of Sciences of Ukraine*
*47 Nauky Ave., Kharkov, 61103, Ukraine*
E-mail: pastur@ilt.kharkov.ua

Spin valves based on materials in which the spin-flip is suppressed by the spatial separation of charge carriers, while maintaining electric neutrality in the valve volume, are considered. The possibility of using these valves as electric batteries is discussed. It is shown that if the potential difference across the valve is controlled, incommensurability effects such as the "devil's staircase" may be expected, which are associated with the Coulomb interaction and redistribution of electrons occurring while the battery is charged and discharged. The effects of the emergence and vanishing of spontaneous spin polarization of conduction electrons with a change in the Fermi level in the valve are predicted. Such spin valves can also be used in implementing spintronic memory cells, supercapacitors, and similar devices.

Key words: spin polarization, spin-flip, spin valve.

## 1. Introduction

Spin valve devices play an important role in modern spintronics. In particular, the effect of giant magnetoresistance arising in locked spin valves is widely employed in technical applications, for example, in state-of-the-art hard disks, magnetic sensors, *etc*. [1–4]. This article considers a special spin valve as a possible version of a spintronic electric (*i.e.*, producing a charge current) rechargeable battery, or supercapacitor, in contrast to "spin" batteries (see, for example, [5,6]) and other recent proposals [7,8], including actively discussed quantum batteries [9–11].

In the simplest case, a spin valve is a sandwich system that consists of a non-magnetic conductor with two adjacent outer plates, which are magnetized conductors, at its opposite surfaces. Since the passage of the conduction electron through a magnetized plate depends on the orientation of the electron spin relative to the magnetic moment of the plate, the conductivity of such a system is determined by their relative orientation. The minimum conductivity is attained if the magnetic moments of the plates are oriented in opposite directions, in the so-called "locked" spin valve, in which the well-known giant magnetoresistance effect is exhibited (see the Nobel lectures [13,14] and references quoted there). It is of importance to note that a difference (splitting) of the spin concentrations occurs in such a valve in the presence of a potential difference, since a spin oriented in one direction enters the nonmagnetic region through one of the plates, while an oppositely oriented spin escapes from it through the other plate, so that the so-called spin accumulation takes place [15–17], resulting in ideal conditions for blocking of the current through the valve. It should be noted that since the electron is both a spin carrier and a charge carrier, a locked spin valve can perform as an electric capacitor and even a rechargeable battery, featuring certain advantages over conventional electrochemical rechargeable batteries (see the Conclusion). On the other hand, a serious adverse feature is the possible spin relaxation (spin-flip), both in the magnetized plates and in the nonmagnetic part. As a result, the locked valve nevertheless conducts current even if completely polarized magnetic plates are used. This feature may significantly limit the options of using a locked valve as an electric rechargeable battery.

If the spin-flip inside the valve is neglected, the principle of battery operation is as follows. The splitting of spin concentrations, accumulated by externally charging the potential difference, becomes thermodynamically nonequilibrium if the valve is connected to a circuit without a charging voltage and, therefore, recuperates the charge current back to the circuit like an ordinary electric battery. In the absence of a spin-flip, the electrons with a higher spin concentration leave the valve due to diffusion, while the electrons with a lower spin concentration penetrate the valve. In a locked valve, the magnetized plates that form that valve, ensure that the electrons with a certain spin are drawn in from only one side through the corresponding plate, and those with the opposite spin are pushed out on the opposite side through the other plate. When the magnetized plates violate the symmetry of diffusion of electrons with different spin directions, it creates a charge current in the system. In other words, the relaxation of the spin-nonequilibrium valve state in the electric circuit is accompanied by the electric current.

A model of an ideal spin valve, which consists of half-metal* magnetic plates with a non-magnetic conductor sandwiched between them in the absence of a spin-flip, is considered in Section 2. It should be noted that this model can also be applied in the presence of a spin-flip; however, only at high frequencies and for microscopic dimensions, when the characteristic spin-flip times significantly exceed the valve operating times.

Section 3 presents an example of "layered" antiferromagnetic (meta) materials in which a spin-flip is not possible, due to the spatial separation of carriers with two spin directions. The synthesis of such structures would allow the creation of highly efficient storages of energy and charge, whose implementation is not limited to high frequencies and small dimensions. Also considered is an essential feature of the spin valve made of the proposed layered materials – incommensurability effects like the "devil's staircase", which are a direct manifestation of the complexity of the charging process that occur in such rechargeable batteries.

Section 4 shows the well-pronounced discrete nature of the "devil's staircase" steps that can be used to develop spintronic memory cells; qualitative theoretical considerations are confirmed by numerical results.

* Half-metal conductivity means that the density of states at the Fermi level is non-zero for only one of two possible directions of the conduction electron spins.

## 2. Ideal spin valve

We now consider the ideal locked spin valve, which consists of half-metal magnetic plates H with oppositely directed magnetic moments, and a non-magnetic normal conductor N in the absence of a spin-flip (Fig. 1). The magnetization of the plates can be either intrinsic or induced by an external magnetic field. It sets the directions of the conduction electron spins in the plates, which we denote by arrows ↑ and ↓. The N-conductor does not have exchange splitting of the spin subbands, so that the densities of states of ↑-electrons and ↓-electrons at the Fermi level are the same. On the other hand, the exchange spin splitting in the magnetic plates H is assumed to be so strong that at the Fermi level only one of the two densities of states is other than zero (only ↑ or ↓, depending on H magnetization directions). We assume that the temperature is $T = 0$. Therefore, ↑-electrons can pass through the left H plate, while ↓-electrons can pass through only the right plate. If a potential difference is created between the plates, the electrons with spins oriented in one direction start entering the conductor N, while the opposite-spin electrons go away. This process generates the spin splitting δη of chemical potentials. For definiteness, we assume that the electrochemical potential of the left plate is a higher than that of the right plate. In equilibrium, the currents cease flowing, which leads to the constancy of the corresponding electrochemical potentials.

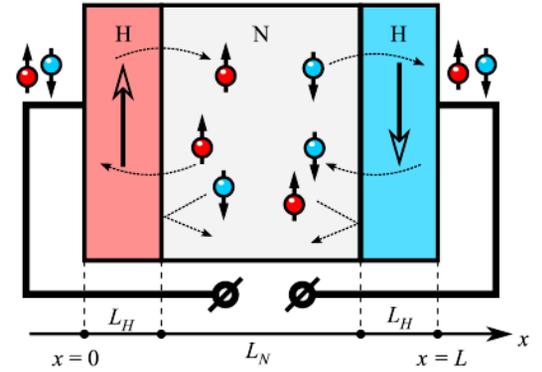

*Fig. 1.* Ideal spin valve N is the non-magnetic conductor, H are half-metal conductors (plates). Electrons with a spin are shown as circles with arrows ↑,↓. It is assumed that the left H-plate gives a pass to spin-up electrons (↑), and the right one to spin-down electrons (↓).

Let $\mu_\sigma$, $\sigma = \uparrow,\downarrow$, be the electrochemical potentials of the spin components, φ the electric potential, $\eta_\sigma = \mu_\sigma - e\varphi/2$ the chemical potentials of spin components, and $e$ the elementary charge. The $x$-coordinate is measured along the valve axis, so that $x = 0$ corresponds to the left HN interface and $x = L$ to the right NH interface. The thicknesses of the N and H layers are denoted as $L_A$, $A = N,H$ (Fig. 1). The potential difference across the valve

$$V = \varphi(0) - \varphi(L) = \frac{[\mu_\sigma(0) - \mu_\sigma(L)]}{e}, \sigma = \uparrow,\downarrow. \quad (1)$$

Since only electrons with a certain spin pass through the plates, electrons with spin ↑ enter N from the left and are accumulated (since they cannot exit through the right layer H) until $\eta_\uparrow$ is leveled out in N by $e\varphi(0)$, while electrons with spin ↓ go right also until $\eta_\downarrow$ is leveled out in N by $e\varphi(L)$, i.e. until spin diffusion across the boundaries $x = L_H$ and $x = L_H + L_N$ balances the electric "pressure". As a result, similar to the emergence of the usual contact potential difference, a magnetic-contact potential difference emerges at these boundaries, i.e. double electrical layers that provide a jump of φ and $\eta_\sigma$. At the same time, the continuity (and in this case, constancy) of the component $\mu_\sigma$ is restored; see, e.g. [18–20]. The general relationship between the electric potential, the spin components of the chemical and electrochemical potentials, and the densities of states in H and N can be obtained from the Poisson equation with the charge density derived from the quasi-equilibrium single-particle distribution. At distances from the boundaries greater than their screening radius $r$, the condition of electric neutrality yields a constant electrochemical potential for spins ↑ on the HN interface $x = L_H$ [19,21]:

$$\eta_\sigma(x) = \frac{\Pi_\sigma(\varepsilon_F, x)}{\Pi_\uparrow(\varepsilon_F, x) + \Pi_\downarrow(\varepsilon_F, x)}[\mu_\sigma(x) - \mu_{-\sigma}(x)] \quad (2)$$

$$\varphi(x) = \frac{1}{e}\frac{\Pi_\uparrow(\varepsilon_F, x)\mu_\uparrow(x) + \Pi_\downarrow(\varepsilon_F, x)\mu_\downarrow(x)}{\Pi_\uparrow(\varepsilon_F, x) + \Pi_\downarrow(\varepsilon_F, x)}, \quad (3)$$

$$\Pi_\sigma(\varepsilon, x) = \begin{cases} \Pi_{\sigma N}(\varepsilon), & L_H < x < L_H + L_N; \\ \Pi_{\sigma H}(\varepsilon), & L_H < x, 2L_H + L_N > x > L_H + L_N, \end{cases} \quad (4)$$

where $\Pi_{\sigma A}(\varepsilon)$, $\sigma = \uparrow, \downarrow$, $A = H, N$ is the density of states of $\sigma$-electrons in plate H or non-magnetic conductor N, respectively, and $\varepsilon_F$ is the Fermi energy. These formulas show that $\Pi_\sigma(\varepsilon_F, x)$, $\varphi(x)$, and $\eta_\sigma(x)$ undergo a jump at the NH interface $x = L - L_H$, which leads to the emergence of additional resistance across it [18]. A similar situation is observed at the HN interface $x = L_H$. We have $\mu_\sigma = \text{const}$ at the boundaries where the transition of $\sigma$-electrons is possible. (Further discussion is presented in the Appendix.)

The equilibrium potentials of a locked ideal spin valve are schematically shown in Fig. 2. As can be seen, the diffusion energy accumulated due to the spin splitting of chemical potentials compensates for the applied difference of electric potentials. Consequently, the energy stored in the ideal locked spin valve per unit volume of the N-conductor, at temperatures low compared with the Fermi energy $\varepsilon_F$ and at $eV < \varepsilon_F$, has the form

$$\delta\mathcal{E} \approx \int_0^\infty \varepsilon[n_\uparrow(\varepsilon) + n_\downarrow(\varepsilon) - 2n(\varepsilon)]\Pi(\varepsilon)d\varepsilon \sim e^2 V^2 \Pi_N, \qquad (5)$$

where $n$ is the equilibrium (Fermi) distribution and $\Pi_N$ is the density of states at the Fermi level in N. Thus, if $\Omega_N = L_N^3$ is the N-conductor volume, the electric capacitance of the locked spin valve is

$$C \sim \frac{\Omega_N \delta\mathcal{E}}{V^2} \sim e^2 \Pi_N \Omega_N \sim \frac{me^2 \sqrt{m\varepsilon_F}}{\hbar^3} \Omega_N \sim$$

$$\sim \frac{me^2}{\hbar^3} \frac{1}{\lambda_F} \Omega_N \sim \frac{\Omega_N}{\lambda_F^2} \sim L_N \frac{L_N^2}{\lambda_F^2}, \qquad (6)$$

where it is taken into account that for normal metals, the Coulomb energy at the Fermi wavelength $\lambda_F$ is of the same order as $\varepsilon_F$. This shows that the effective "capacity" contains a large parameter $(L_N/\lambda_F)^2$. The ideal locked spin valve, in terms of energy storage, would be an ideal supercapacitor or a rechargeable battery that does not use any chemical reactions.

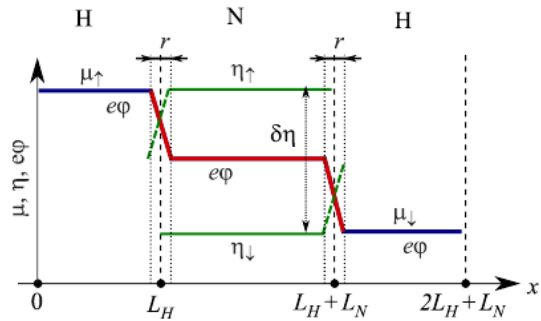

*Fig. 2.* Electrochemical, chemical, and electric potentials in the H–N–H locked ideal spin valve with the potential difference $V$ applied to the H-plates. The difference between the spin components of chemical potentials is proportional in equilibrium to the applied potential difference, $\delta\eta = eV$. A potential jump (the magnetic-contact potential difference) is observed at the HN and NH boundaries in the double electric layer, at a distance from the boundaries of the order of the screening radius $r$.

Of course, if currently known metals and their alloys are discussed, then even if ideal half-metals are used as plates and temperatures are low, the problem of conduction-electron spin-flip in a non-magnetic metal persists, which destroys the described ideal picture. The electrons that experienced the spin-flip apparently begin leaking through the half-metal, which leads to the emergence of a potential gradient and the corresponding leakage current. Therefore, in the case of ordinary metals or semiconductors, such an effect can manifest itself only in the form of reactance at frequencies higher than the inverse spin-flip frequency $\tau_{sf}^{-1}$, and only in microscopic structures smaller than the spin-flip diffusion length $\lambda \sim \sqrt{lv_F \tau_{sf}}$, where $l$ is the mean free path of conduction electrons and $v_F$ is the Fermi velocity. For example, in the case of copper under the experimental conditions of Ref. [22], $\lambda \sim 1$ μm and the corresponding frequency $\tau_{sf}^{-1} \sim 10^{-2} - 10^{-1} \text{ps}^{-1} \sim 10^{10} - 10^{11}$ Hz is rather high. Consequently, in order for low-frequency effects, and even more so energy accumulation to be manifested, a completely new material is needed where there is no spin-flip.

## 3. Metamaterial without spin-flip and the "devil's staircase"

To realize an ideal spin valve, we need materials that are not electrically different from a conventional conductor, but in which a spin-flip of conduction electrons is impossible. Apparently, topological dielectrics [23–25], in particular, layered topological dielectrics with chiral boundary conditions [26] may be used for this purpose. Another option is a man-made composite material — a layered structure (conductive quasi-two-dimensional layers or quasi-one-dimensional fibers), in which the conductive layers are located close enough (at a distance $d$ smaller than the screening radius $r$) to ensure the electrical uniformity of the plate, but are sufficiently separated to suppress electron tunneling between the layers. Electrons can only move in such structures within their layer. The layers themselves should consist of magnetic atoms that are magnetized oppositely with respect to each other, as shown in Fig. 3. From the viewpoint of the bands, such layers must be actual low-dimensional half-metals, similar to those described in [27–30].

Figure 3 shows that electrons in such media carry current along the layers, as in an ordinary nonmagnetic conductor, but the spin-flip in them would be associated with the transition between layers, due to the absence of free spin subbands of the opposite spin direction in each such layer. Therefore, the spin-flip will be suppressed due to spatial separation, similar to the suppression of electron-hole layer recombination considered in studies of electron-hole pairing [31–37]. A significant difference between the proposed layered material and the usual non-magnetic conductor is that if a nonequilibrium electron distribution between the layers emerges in such a material, it will relax only by decreasing the Coulomb energy of the interlayer interaction, since spin-flips are forbidden.*

---

* The spin-flip at the HN and NH boundaries due to some combined scattering apparently should also be suppressed, such as by breaking the contact between the corresponding chains at the interfaces, as shown in Fig. 3.

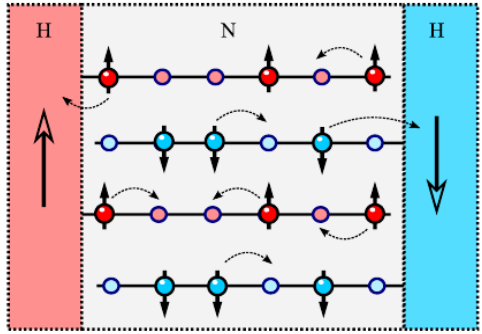

*Fig. 3.* Diagram of a layered N-material. No tunneling occurs between the layers; charge and spin carriers move only within the layers.

If the proposed material is used as a non-magnetic conductor N in a locked spin valve with a potential difference $V$, the number of electrons with spin $\uparrow$ increases in the layers with the corresponding direction of magnetization, while in oppositely magnetized layers the number of $\downarrow$ electrons decreases. The difference in chemical potentials $\delta\eta = eV$, both between the layers and in the total volume, is apparently fixed by the applied voltage. The charges are redistributed due to the Coulomb interaction, and the spin degree of freedom of the electrons will be significantly entangled with their charge degree of freedom. However, in general the system remains electrically neutral ($d \ll r$), and the distribution of potentials corresponds to that displayed in Fig. 2.

We consider now the case of a narrow-band-gap non-magnetic metal, such that the width $t$ of its conduction band is less than a typical change in the electron energy when hopping to a neighboring site $\sim (a/R)E$ (here $a$ is the distance between the conductor lattice nodes, $R$ is the average distance between electrons, and $E$ is the average interaction energy per electron). As shown in Ref. [38], a macroscopic state, referred to as the "frozen electronic phase" (FEP), is formed in this case in the conductor. This state emerges due to a combination of the long-range repulsive potential and discrete nature of the narrow-band-gap electron dynamics (*i.e.*, electrons move along the conductor by hopping between the nearest nodes of the conductor lattice with equivalent atomic orbitals). The Bloch states are completely destroyed in this situation, and the electrons are localized in atomic-size quantum traps. Due to the dynamic nature of the localization, the heating of the system cannot release electrons from the traps, and, therefore, the FEP, unlike the Wigner crystal [39], does not transform into the Fermi liquid, over a rather wide temperature range. The emergence of the FEP makes it possible to describe the system under study within the classical lattice model called the "generalized Wigner crystal" [40–46].

The Hamiltonian of this model has the form

$$\mathcal{H} = \frac{1}{2}\sum_{i,i'}\sum_{k,k'} U_{k,k'}^{i,i'} n_k^i n_{k'}^{i'} - \mu_\uparrow \sum_{i,k} n_k^{2i} - \mu_\downarrow \sum_{i,k} n_k^{2i+1}, \quad (7)$$

where the superscript numbers the layers, the subscript indicates the position in the layer, $n_k^i = 0, 1$ are the occupation numbers of electrons at site number $k$ in the layer $i$, and $U_{k,k'}^{i,i'}$ is the repulsion potential that operates between the electrons at sites $k$ and $k'$, located in layers $i$ and $i'$, respectively. It is assumed that even layers have spin polarization $\uparrow$ and the odd ones $\downarrow$, *i.e.* the spin index is the same as the layer index.

It should be noted that, since the distance between the layers is much smaller than the screening radius, the interaction between all electrons in various layers should be taken into account, as a result of which the Coulomb potential of the layer cannot be set as a thermodynamic quantity. Instead, the charges of all layers "screen" each other, establishing a common electric potential in N. The electron density in the layers is fixed by the chemical potential of the plates. Namely, at equilibrium the electrochemical potential $\mu_\sigma$, $\sigma = \uparrow, \downarrow$ of each layer, which corresponds to a single spin component, is imposed by a plate with the same spin momentum; these potentials are the same in even and odd layers and determine $\mu_\sigma$ and $\eta_\sigma$ of the entire conductor N. Since the final state of a charged battery is of interest, we consider the problem of the equilibrium state of chains for a given splitting of chemical potentials $\delta\eta = eV$. The processes of charging and discharging *per se* are determined by the kinetic term of the Hamiltonian, which is neglected in comparison with the interaction energy, *i.e.* these processes are assumed to be adiabatic. Therefore, this model and the scenario below can be used only for times that are much longer than the charging time of such a battery and, even more so, the energy relaxation time. Non-stationary phenomena that occur in the process of charging and discharging require separate consideration.

The properties of a generalized Wigner crystal in the one-dimensional case have been well studied [40–43,46,47]. Although the employed model of a narrow-band-gap nonmagnetic metal N cannot be applied to this case, because the transverse layers have indices that are tightly connected to spin, we recall the results of this study. It is based on the Hamiltonian (*cf.* (7))

$$\mathcal{H} = \frac{1}{2}\sum_{k,k'} U_{k,k'} n_k n_{k'} - \mu \sum_k n_k. \quad (8)$$

Here and below, all energies (and temperature $T$) are calculated in units of the Coulomb energy $e^2/\varepsilon a$, where $e$ is the electron charge, $a$ is the lattice period, $\varepsilon$ is the dielectric constant of the medium, and all distances are measured in units of $a$.

If the potential $U_{k,k'} = U(|k - k'|)$ is
1) monotonically decreasing;
2) convex everywhere; and
3) $U(r) \sim r^{-1-\delta}, \ r \gg 1, \ \delta > 0$, \quad (9)

then, at $T = 0$, the dependence of the particle concentration

$$c = \frac{1}{N}\sum_{k=1}^{N} n_k$$

on chemical potential $\mu$ is described by a resembling the Cantor staircase function (Lebesgue–Cantor-type function), *i.e.* represents a fractal curve like the "devil's staircase" [42]. Each step in this dependence corresponds to an electronic crystal with particle concentration $c = p/q$, where $p$ and $q$ are integers. Such a crystal has a period (in units of the distance $a$ between the conductor's lattice sites) equal to $q$ and contains $p$ particles per unit cell. The coordinates $r_j$ of the particles of this electron crystal are described by the formula [41]

$$r_j = [j/c + \phi]. \qquad (10)$$

Here the index $j = 0, \pm1, \pm2, ...$ numbers the particles, the symbol $[...]$ denotes the integer part of the number, and the real initial phase $\phi$ is determined by the choice of the origin.

The width of the "devil's staircase" step (*i.e.*, the stability boundary of the electronic crystal with concentration $c = p/q$) is determined by the formula [42]

$$\Delta\mu\left(c = \frac{p}{q}\right) = p\sum_{l=1}^{\infty} l[U(lq + 1) + U(lq - 1) - 2U(lq)]. \qquad (11)$$

It should be noted that if conditions (9) are fulfilled, the properties of the one-dimensional generalized Wigner crystal at $T = 0$ do not depend in qualitative terms on the potential $U$. Its specific form affects only the widths of the "devil's staircase" steps (11), while the ground state structures corresponding to these steps are independent of $U$ (see (10)).

The fractal dependence of the concentration on the chemical potential is destroyed in real physical systems, due to such factors as temperature, finiteness of the radius of interaction between particles $R_0$, and the presence of defects and impurities. All steps with widths $\Delta\mu \ll T$ vanish at any finite temperature, and the remaining steps are smoothed out. The finiteness of $R_0$ results in the destruction of electronic crystals whose period $q \gg R$. In other words, the infinite set of self-similarity scales inherent in fractal dependences vanishes, but a finite set of scales at which fractal dependence occurs, persists. Nevertheless, a number of non-trivial thermodynamic properties of the one-dimensional generalized Wigner crystal does persist in these cases (see Refs. [43,48]). For example, if the interaction only between the nearest particles (but not between the nearest lattice sites!) is taken into account, depending on $c(\mu)$, the "devil's staircase" steps persist, and correspond to concentrations of the form $c = 1/q$, $q = 1, 2, ...$ If interaction between the particles next to the nearest neighbors is taken into account, additional steps emerge that correspond to concentrations $c = 2/q$, *etc.*

The disordered character of lattice site positions or the presence of impurities also destroys the "devil's staircase."If the disorder is very strong, electronic crystals are apparently fully destroyed, and the fractal dependence of the concentration on the chemical potential vanishes. In the weak disorder region, electronic crystals are divided into blocks of random sizes. A typical size of such blocks is $\sim 1/D^2$, where $D$ is the variance of random positions of lattice sites [49–51]. The positions of the electrons within each block are described as before by Eq. (10), but in this case, the phase $\varphi$ depends on the block number.

The structure of the ground state of a two-dimensional generalized Wigner crystal was studied in Ref. [45]. It is shown that an effective decrease in dimensionality occurs in this case: the ground state is described by the "one-dimensional" formula (10), but the structural elements in this equation are bands – one-dimensional periodic electronic structures – rather than electrons. The shape of these structures is determined by the electron density $c$, their interaction potential $U$, and the lattice geometry.

Studying the properties of quasi-two-dimensional generalized Wigner crystals is a challenging task [44,45]; therefore, it seems reasonable to start the discussion with a simple non-trivial model that consists of two chains. It should be noted that conditions 1 and 2 in Eq. (9) for the potential of electron-electron interaction are fulfilled automatically. The fulfillment of the screening condition (requirement 3 in Eq. (9)) in the case under consideration is ensured by the magnetic plates that limit these chains. We assume that spins are directed upward ($\uparrow$) in one of the chains, and downward ($\downarrow$) in the other. The external parameter is in this case the difference between electrochemical potentials $\mu_\uparrow = \mu_\downarrow$, which (see (1)) is determined by the external potential difference $V$ applied to the spin valve. The Hamiltonian of this model has the form

$$\mathcal{H} = \frac{1}{2}\sum_{\sigma,\sigma'=\uparrow,\downarrow}\sum_{k,k'} U_{k,k'}^{\sigma,\sigma'} n_k^\sigma n_{k'}^{\sigma'} - \mu_\uparrow \sum_k n_k^\uparrow - \mu_\downarrow \sum_k n_k^\downarrow, \qquad (12)$$

where $\sigma = \uparrow, \downarrow$ are the chain indices, and $n_k^\sigma = 0, 1$ are corresponding occupation numbers.

In the considered model, the external potential difference $V$ fixes the difference of chemical potentials $\mu_\uparrow$ and $\mu_\downarrow$, which imposes certain electron densities in the chains with spin $\uparrow$ and $\downarrow$. It is shown below that this model cannot be reduced to a one-dimensional generalized Wigner crystal even in the case of non-interacting chains [40–43,46].

We now introduce the variables $n^\pm = (n^\uparrow \pm n^\downarrow)/2$ and assume that the chains are identical:

$$U_{k,k'}^{\uparrow,\uparrow} = U_{k,k'}^{\downarrow,\downarrow} = U_{k,k'}^{(0)}, \quad U_{k,k'}^{\uparrow,\downarrow} = U_{k,k'}^{\downarrow,\uparrow} = U_{k,k'}^{(1)}.$$

This leads to the Hamiltonian

$$\mathcal{H} = \sum_{k,k'}\left(U_{k,k'}^{(0)} + U_{k,k'}^{(1)}\right) n_k^+ n_{k'}^+ + \sum_{k,k'}\left(U_{k,k'}^{(0)} - U_{k,k'}^{(1)}\right) n_k^- n_{k'}^- \\ - \mu\sum_k n_k^+ - \delta\eta \sum_k n_k^- \qquad (13)$$

with separable variables, where $\mu = \mu_\uparrow + \mu_\downarrow$ is the initial electrochemical potential of the system (independent of the applied potential $V$), and $\delta\eta = \mu_\uparrow - \mu_\downarrow$ is the splitting of the chemical potential of spins $\uparrow$ and $\downarrow$, and, as indicated above, in this spin valve we have $\delta\eta = eV$. This Hamiltonian contains two external parameters that govern individual periodic phases: $\mu$ controls the charge phase, and $V$ or $\delta\eta$ is the spin phase (polarization), but the interaction can entangle these phases. Therefore, the problem cannot be reduced in the general case to a single Ising chain, as in [45,46], since $n^- = -1, 0, 1$ and $n^+ = 0, 1, 2$.

In the physically reasonable case, when the interaction between the layers is small compared with the interaction in the layer $\left(U_{k,k'}^{(0)} \gg U_{k,k'}^{(1)}\right)$, the leading order of parameter $U^{(1)}/U^{(0)} \ll 1$ has an energy minimum, and hence, the main contribution to the partition function at low temperatures is provided by the energy minima of two independent chains ↑ and ↓, owing to which the results for a single chain obtained in [40–43,46] may be used. The chemical potentials of the chains in the case under consideration have the form

$$\mu_\uparrow = \mu + \frac{\delta\eta}{2}, \quad \mu_\downarrow = \mu - \frac{\delta\eta}{2},$$

therefore, according to Ref. [42] (see also above), the concentrations $c_\sigma$, $\sigma = \uparrow, \downarrow$ are Lebesgue – Cantor type functions:

$$c_\uparrow = C\left(\mu + \frac{\delta\eta}{2}\right),$$
$$c_\downarrow = C\left(\mu - \frac{\delta\eta}{2}\right), \quad (14)$$

*i.e.* two different "devil's staircases."

Thus, in the leading approximation with respect to $U^{(1)}/U^{(0)} \ll 1$, the spin valve under consideration consists of two non-interacting one-dimensional electron crystals located in chains ↑ and ↓. The total energy is apparently equal to the sum of the energies of these crystals. According to Eq. (10), the energies of the crystals are degenerate with respect to phases $\phi_{\uparrow,\downarrow}$, where $\phi_{\uparrow,\downarrow}$ are electronic crystal phases of chains ↑ and ↓, respectively. In other words, these energies remain unchanged if the chains are shifted by any distance that is a multiple of the lattice period $a$. If interaction between the chains (*i.e.*, first order in $U^{(1)}/U^{(0)} \ll 1$ is taken into account), this degeneracy is removed, and the total energy is minimized further by varying the phase difference.

Therefore, in the leading approximation, the electron density polarization in a given spin valve (see (14))

$$m \equiv \frac{c_\uparrow - c_\downarrow}{c_\uparrow + c_\downarrow} = \frac{C\left(\mu + \frac{eV}{2}\right) - C\left(\mu - \frac{eV}{2}\right)}{C\left(\mu + \frac{eV}{2}\right) + C\left(\mu - \frac{eV}{2}\right)} \quad (15)$$

is a rational fractional combination of two Lebesgue–Cantor-type functions $C$ and, consequently, has a complex stepwise dependence on $V = \delta\eta/e$ (Fig. 4).

It should be noted that the dependence $m(V)$ determines the number of particles that can enter and exit such a locked spin valve, and, therefore, the EMF of this rechargeable battery.

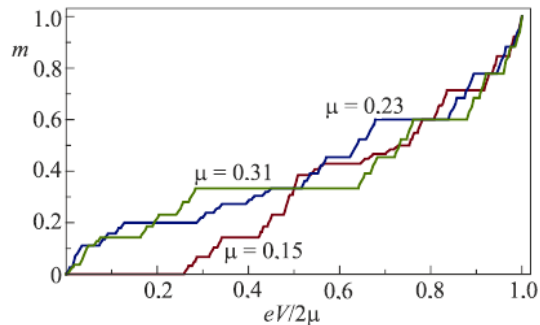

*Fig. 4.* Example of a qualitative dependence of the magnetization $m$ on the potential, represented as a combination of Lebesgue

functions (15) in dimensionless units.

## 4. Numerical results

We consider the case of isotropic interaction at $T = 0$ and choose the potential contained in Eq. (12) in the form

$$U_{k,k'}^{\sigma,\sigma'} = U\left(\left|r_k^\sigma - r_{k'}^{\sigma'}\right|\right) = \left(\left|r_k^\sigma - r_{k'}^{\sigma'}\right|\right)^{-2}. \quad (16)$$

This potential apparently satisfies conditions (9). Recall that the structure of the dependences $m(\mu)$ and $m(\delta\eta)$ (see (15)) is qualitatively independent of $U$ under conditions (9). The dependences of $m$ on $\mu$ at fixed values of $\delta\eta$, and the dependences of $m$ on $\delta\eta$ at fixed values of $\mu$, are displayed in Figs. 5 and 6, respectively. Both dependences were obtained by searching for the energy minimum (12) while testing all the states for a system of 2 chains, each of which contains 12 sites (*i.e.*, by testing $2^{24}$ states).

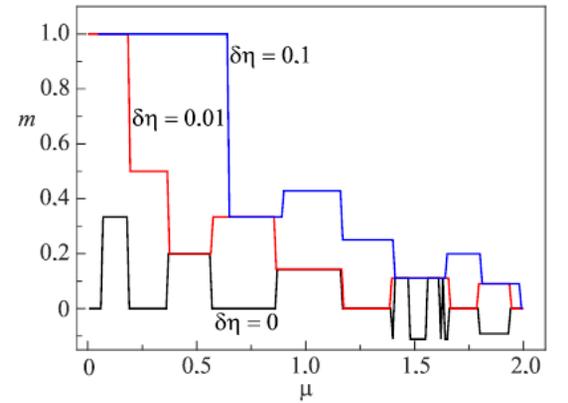

*Fig. 5.* Dependences of the magnetization $m$ on the chemical potential $\mu$ at different values of spin splitting, $\delta\eta$. $T = 0$. The system consists of two chains each containing 12 sites.

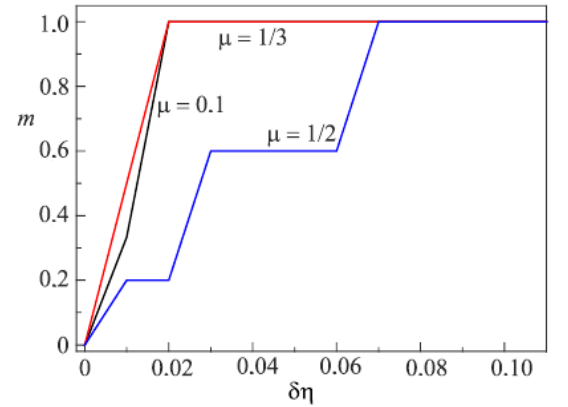

*Fig. 6.* Dependences of the magnetization $m$ on $\delta\eta$ at various values of $\mu$ for a system of 2 chains, each containing 12 sites. $T = 0$. The interaction potential has the form of Eq. (16).

The next step is to explore the effect of non-zero temperature on the properties of the system under study. We consider the case of two isolated chains in the nearest-neighbor approximation. The potential can be represented in this approximation as

$$U_{k,k'}^{\sigma,\sigma'} = \delta_{\sigma,\sigma'} U\left(\left|r_k^\sigma - r_{k'}^{\sigma'}\right|\right) = \delta_{\sigma,\sigma'} U(l), \quad l = \left|r_k^\sigma - r_{k'}^{\sigma'}\right|. \quad (17)$$

The potential $U(l)$ in Eq. (17) is chosen in the form $U(l) = l^{-2}$ (see (16)).

The isothermal partition function of an isolated chain has the form [43]

$$Z = Z(N, \mu, T) = \sum_{\{n_l\}} \exp\left(-\frac{E\{n_l\} + \mu L\{n_l\}}{T}\right) W\{n_l\}, \quad (18)$$

where $E\{n_l\} = \sum_{l=1}^{\infty} lU(n_l)$ is the interaction energy of the electrons of the chain and $W\{n_l\} = N!/\prod_{l=1}^{\infty} n_l!$ is the statistical weight. The notation $\{n_l\}$, $n_l = 0, 1, ...$ corresponds to the configuration (distribution along the chain) of electrons located at distances $n_1, n_2, ...$ from each other. Summation is performed over all configurations that satisfy the condition $N = \sum_{l=1}^{\infty} n_l$. The chain length is $L\{n_l\} = \sum_{l=1}^{\infty} ln_l$. Then

$$Z(N, \mu, T) = \left(\sum_{l=1}^{\infty} \exp\left(-\frac{U(l) + \mu l}{T}\right)\right)^N \quad (19)$$

and the corresponding thermodynamic potential (Gibbs free energy) is

$$\Phi(N, \mu, T) = -TN \ln\left[\sum_{l=1}^{\infty} \exp\left(-\frac{U(l) + \mu l}{T}\right)\right]. \quad (20)$$

The electron density $c$ of chain $\sigma$, as a function of $\mu$ and temperature $T$, has the form

$$c(\mu_\sigma) = \frac{\sum_{l=1}^{\infty} \exp\left(-\frac{U(l) + \mu_\sigma l}{T}\right)}{\sum_{l=1}^{\infty} l \exp\left(-\frac{U(l) + \mu_\sigma l}{T}\right)}, \quad (21)$$

The dependences $c(\mu)$, $m(\mu)$, and $m(\delta\eta)$ at various temperatures $T$ are displayed in Figs. 7, 8, and 9, respectively.

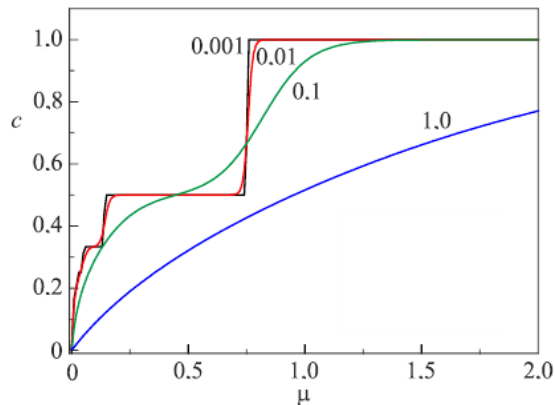

*Fig. 7.* Dependences of the electron density $c$ on a single isolated chain on $\mu$ (see (21)) at various temperatures.

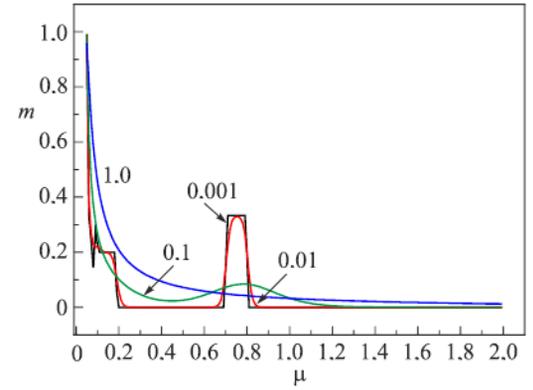

*Fig. 8.* Dependences of the magnetization $m$ of a system of 2 isolated chains on the chemical potential $\mu$ at spin splitting $\delta\eta = 0.1$ and various values of $T$.

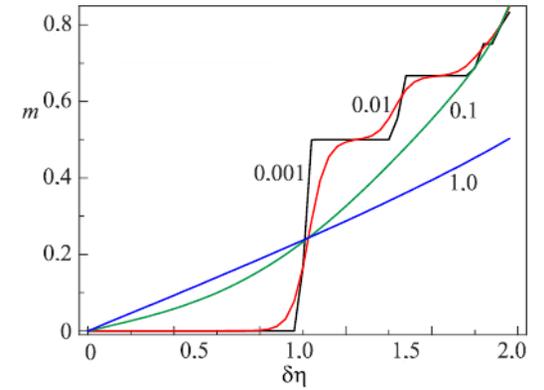

*Fig. 9.* Dependences of the magnetization $m$ of a system of 2 isolated chains on spin splitting $\delta\eta$, at $\mu = 0.5$ and various values of $T$.

As can be seen, the finite temperature blurs the fine structure of the steps, but the qualitatively stepwise dependence persists to $T \lesssim 10^{-2}$. For typical values of $a = 3-5$ Å, we obtain $T \leq 10^{-2} \frac{e^2}{\varepsilon\varepsilon_0 a k_B} \approx 100$ K (here $k_B$ is the Boltzmann constant).

Figures 5 and 8 show the manifestation of an unusual effect that consists of the polarization of the conduction electron density $m$ as a function of not only the natural parameter $\delta\eta$, but also of the overall electrochemical potential $\mu$, *i.e.* actually of the common Fermi level, and therefore, the total charge of electrons. This implies that the emergence of spontaneous spin polarization of electron density, or its vanishing, is energetically favorable when the Fermi level in this system shifts. The landscape of minima and maxima is stepwise and rather involved; see Fig. 5.

## 5. Conclusion

It is shown that under certain conditions, a locked spin valve can perform as a rechargeable battery in which energy is stored in the form of the difference between chemical potentials of the spin components, due to the spin diffusion "resisting" the Coulomb potential difference, and without involvement of any chemical processes. The reservoir of such a rechargeable battery is a conductor without exchange splitting and with a suppressed spin-flip, which is located between magnetic plates with exchange splitting to half-metal conductivity. To realize such a rechargeable battery, new materials are needed in which the spin-flip of conduction electrons would be suppressed to the maximum possible extent. A mechanism is proposed for such suppression, based on the spatial separation of electrons with differently directed spins in low-dimensional narrow-gap conductors. It is shown that effects such as a combination of spin polarization "devil's staircases" can be realized in these conductors at low density and weak screening of charges. The parameters of the staircases are the Fermi energy and the applied potential difference. Numerical results revealed a number of interesting features of such a rechargeable battery, in particular, the emergence or vanishing of spontaneous spin polarization of conduction electron density, if the electrochemical potential changes. The obtained stepwise dependence of the chemical potential spin splitting (and, hence, the EMF of this type of storage) on the external voltage may also be of interest from the perspective of developing memory cells.

It should be noted that modern technology enables the creation of spin valves and quantum rechargeable batteries of only a very small (micron) sizes. The capacities of such rechargeable batteries are, as a result, also extremely limited. For example, the energy stored in quantum batteries that are currently under active discussion is of the order of $10^{-3}$ eV [12].

As follows from Eq. (5), in the considered model the energy accumulated by the rechargeable battery with volume $\Omega$ at equilibrium with a charging voltage $V$ is, by order of magnitude, equal to

$$\mathcal{E}_\Omega \sim \frac{\Omega}{\lambda_F^3} \frac{e^2 V^2}{\varepsilon_F}.$$

In metals $\lambda_F \sim 0.1$ nm $= 10^{-10}$ m. If the rechargeable battery volume is of the order of one cubic micron, we obtain

$$\mathcal{E}_{\Omega \sim 1\,\mu m^3} \sim \left(\frac{10^{-6}}{10^{-10}}\right)^3 \frac{e^2 V^2}{\varepsilon_F} \sim 10^{12} \frac{e^2 V^2}{\varepsilon_F}.$$

At a voltage typical of this type of systems, $V = 10$ mV, and $\varepsilon_F$, *e.g.* ~ 1 eV, the battery capacity is

$$W = \frac{1}{3600} \frac{\mathcal{E}_{\Omega \sim 1\,\mu m^3}}{V} \sim 10^{-16} \text{A} \cdot \text{h}.$$

It is of importance to note that if the rechargeable battery size is enlarged to centimeters, and voltage is increased to 1 V, a rechargeable battery could be created with energy storage $\mathcal{E}_\Omega \sim 10^{26} \frac{e^2 V^2}{\varepsilon_F}$, a value whose order of magnitude corresponds to the capacity of a laptop battery.

It should be noted that one of the advantages of the proposed rechargeable battery is that the energy needed to generate charge current is accumulated in the volume (*i.e.*, in the spin components of the chemical potential of conduction electrons) rather than on the surface of the plates. In addition, unlike conventional batteries, a spin battery is not associated with chemical reactions, which involves the inhomogeneity of reagent recovery that sooner or later leads to the deterioration of the battery and, eventually, its failure.

We also note that, since the capacitance of a conventional capacitor as a charge reservoir is determined by the surface area of the plates on which the charge is directly stored, increasing its capacity by increasing the number of layers and reducing their thickness (up to nanoscale values), while maintaining a limited volume, hinders uniform contact between large surfaces and, thus, increases the probability of breakdown. In addition, the Coulomb interaction between narrow-band-gap conductors sets significant restrictions on the permissible potential difference, which does not cause such a breakdown. In the case studied, the storage reservoir is a bulk conductor, which is located between the magnetic plates and electrically neutral. Even if it is designed as a set of narrow-band-gap conducting channels, the problem of Coulomb breakdown does not occur.

### Appendix. Magnetic-contact potential difference

We consider the contact of the normal metal N ($\Pi_\uparrow = \Pi_\downarrow$) and the magnetic metal M ($\Pi_\uparrow \neq \Pi_\downarrow$, the limiting case of which is the half-metal H) and assume that there is no spin-flip ($\tau_{sf} = \infty$). We assume also that a potential difference is applied to such an MN block. In the first instant, a gradient of the electric potential, which generates currents, is established throughout the entire block. In the diffusion, *i.e.* the ohmic, transport mode, the relation between the current and the potential gradient is local; in addition, the current is proportional to the density of states of the corresponding carriers. Therefore, at the initial moment of time, the currents $j_\uparrow \neq j_\downarrow$ flow in the MN-border from M, while the currents $j_\uparrow = j_\downarrow$ flow out the boundary to N. As shown in Fig. 10a, such an imbalance of currents leads to the splitting of chemical potentials $\mu_\sigma, \sigma = \uparrow, \downarrow$, *i.e.* to spin accumulation, as a result of which the corresponding currents are balanced. It should be noted that the spin splitting exactly balances just the components of the total electrochemical potential $\mu_\sigma$, as a result of which the currents at the boundary are balanced. Since $\mu_\sigma$ is the sum of two quantities $e\varphi$ and $\eta_\sigma$, the continuity condition of the electrochemical potential is not sufficient for simultaneous balancing of $\varphi$ (*i.e.* charge density) and $\eta_\sigma$ (spin densities) with jump-like changes to $\Pi_\sigma$ at the MN interface (Fig. 10b, *cf.* (2), (3)). Thus, in addition to the spin splitting, electrons accumulate on one side of the MN interface and a deficiency of the electrons occurs on the other side (actually "holes" are created). This leads not only to spin splitting, but also to the contact potential difference and, accordingly, the intrinsic resistance of the MN interface. The intrinsic resistance of such contacts and the jump $\varphi$ were calculated in detail in Ref. [18,20] with account of spin-flip processes, but without the simple qualitative explanation presented in this paper. In the limiting case when a half-metal is used as M, one of the spin densities, for example $\Pi_\downarrow$, vanishes, $\mu_\downarrow$ exists only in N, we have $\mu_\uparrow = e\varphi$ in H, and the splitting should stop the current $\downarrow$ in N (see Fig. 11).

It is most obvious in this limiting case why $\varphi$ cannot be continuous at the interface of the normal and magnetic conductor with $\Pi_\uparrow \neq \Pi_\downarrow$. Figure 11 enables easy comprehension that in the absence of a spin-flip, the two contacts, HN and NH, of the locked spin valve completely block the current and form the scenario of potentials shown in Fig. 2.

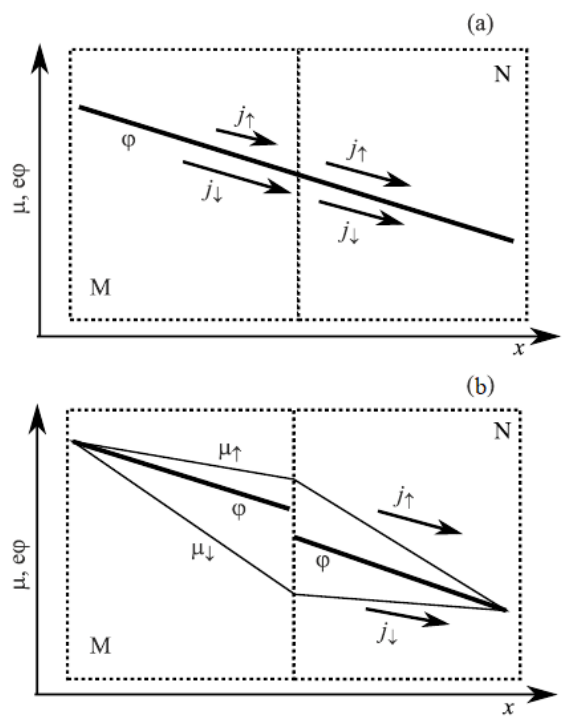

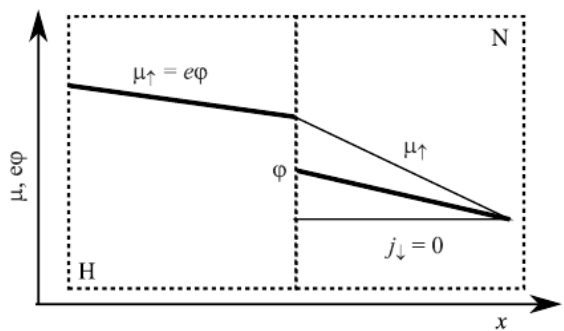

*Fig. 10.* Splitting of electrochemical potentials and the magnetic-contact difference: (a) imbalance at the interface; (b) splitting restored balance.

*Fig. 11.* Potential jump in the H–N case.

Translated by AIP Author Services